\documentclass[aps]{revtex4}
\usepackage{amsmath}
\usepackage{amsmath,epsf}

\begin{document}

\title{Reentrant phase transition of Born-Infeld-dilaton black holes }
\author{Seyed Hossein Hendi$^{1,2}$\footnote{%
email address: hendi@shirazu.ac.ir} and Mehrab Momennia$^{1}$\footnote{%
email address: m.momennia@shirazu.ac.ir}} \affiliation{$^1$
Physics Department and Biruni Observatory, College of Sciences,
Shiraz
University, Shiraz 71454, Iran\\
$^2$ Research Institute for Astronomy and Astrophysics of Maragha (RIAAM),
P.O. Box 55134-441, Maragha, Iran}

\begin{abstract}
We explore a novel reentrant phase transition of four-dimensional
Born-Infeld-dilaton black hole in which the first order phase transition
modify into a zeroth order phase transition below the critical point.
Working in the extended phase space with regarding the cosmological constant
as a pressure, we study the reentrant behavior of phase transition in the
canonical ensemble. We show that these black holes enjoy a zeroth order
intermediate-small black hole phase transition as well as a first order
phase transition between small and large black holes for a narrow range of
temperatures and pressures. We also find that the standard first order
small-large black hole phase transition can modify into a zeroth order type.
This zeroth order phase transition stands between the critical point and the
first order phase transition region. We discuss the significant effect of
the scalar field (dilaton) on the mentioned interesting treatment.
\end{abstract}

\maketitle


\section{Introduction}

The phase transition is one of the interesting topics in thermodynamical
systems. During the phase transition, we often encounter with discontinuity
of a certain thermodynamic property. In order to classify the phase
transition systematically, we can concentrate on the Gibbs thermodynamic
potential and its first and higher derivatives. In the paradigm of the phase
transition, one may use some analogies and symmetries between usual
thermodynamic systems and black holes to deepen his insight into the black
hole physics. It was shown that black holes can be considered as
thermodynamical systems \cite{Davies,Davies1989,Wald} with typical
temperature \cite{Hawking} and entropy \cite{Bekenstein} which satisfy the
first law of black hole thermodynamics \cite{Bardeen}. It has also been
found that they can behave like ordinary thermodynamical systems and undergo
the phase transition \cite{HP}. A more interesting case was observed when
one considers a correspondence of variables $\left( Q,\Phi \right)
\leftrightarrow \left( P,V\right) $\ for charged black holes which results
into the van der Waals like phase transition \cite%
{Chamblin,ChamblinEmparan,Shen}. In recent years, the idea of considering
the cosmological constant as thermodynamical pressure has attracted a lot of
attention in black hole thermodynamics \cite{Caldarelli,Kastor}%
\begin{equation}
P=-\frac{\Lambda }{8\pi },  \label{P2}
\end{equation}%
where the thermodynamical quantity conjugate to $P$\ is the thermodynamical
volume
\begin{equation}
V=\left( \frac{\partial M}{\partial P}\right) _{rep},
\end{equation}%
in which \textquotedblleft $rep$\textquotedblright\ stands for
\textquotedblleft residual~extensive\ parameters\textquotedblright . The
motivation comes from the fact that in some fundamental theories there are
several physical constants, such as Yukawa coupling, gauge coupling
constants, and Newton's constant that are not fixed values. On the other
hand, the cosmological constant stands by pressure side in
Tolman--Oppenheimer--Volkoff equation which shows the cosmological constant
can be considered as thermodynamical pressure. Considering $\Lambda $\ as
the pressure of system leads to a van der Waals like small-large black holes
(SBH-LBH) phase transition which has been investigated by so many authors
(for instance see incomplete list \cite%
{KubiznakMann,Banerjee,Mirza,Mo,Zou,Xu,HendiFaizal,Mandal,HendiPV,RainbowYM,MassiveYM}
and references therein).

The reentrant phase transition (RPT) can occur in a thermodynamical system
whenever a monotonic change of any thermodynamic variable leads into more
than one phase transition so that the final state is macroscopically similar
to the initial state. In our black holes case study, there is a specific
range of temperatures such that black holes undergo a large-small-large
phase transition by a monotonic changing of the pressure. This interesting
phenomenon has been first observed in a nicotine-water mixture \cite{Hudson}%
, and then seen in multicomponent fluids, binary gases, liquid crystals, and
other diverse systems \cite{Narayanan}. In the context of black holes, the
RPT has been observed for Born-Infeld adS black holes \cite{BIadS}, rotating
adS black holes \cite{rotatingadS,rotatingAmin}, dS black holes \cite%
{deSitter}, hairy black holes \cite{hairy}, and adS black holes in massive
gravity \cite{dRGTmassive,massive}. The van der Waals like phase transition
of SBH-LBH in dilaton gravity has been investigated for charged adS black
holes \cite{Zhao}, and also, different types of nonlinear electrodynamics,
such as power Maxwell invariant \cite{PMIdMo,PMId}, exponential \cite{ENED},
and Born-Infeld \cite{PVdilatonH,PVdilatonSh}. More recently, a study
regarding zeroth order SBH-LBH phase transition of charged dilaton black
holes has been done \cite{Dehyadegari}. The purpose of this paper is
studying the thermodynamics of $4$-dimensional Born-Infeld-dilaton black
holes and investigating the RPT in the canonical ensemble of extended phase
space. In addition, we are going to show that this kind of black holes can
undergo a zeroth order SBH-LBH phase transition.


\section{review of solutions and thermodynamics \label{FE}}

Topological Born-Infeld-dilaton black holes in $4$-dimensional spacetime
were constructed by Sheykhi \cite{Sheykhi}. In what follows we concentrate
our attention on the spherical symmetric black holes with negative
cosmological constant. The line element reads%
\begin{equation}
ds^{2}=-f(r)dt^{2}+f^{-1}(r)dr^{2}+r^{2}R^{2}(r)\left( d\theta ^{2}+\sin
^{2}\theta d\varphi ^{2}\right) ,  \label{metric}
\end{equation}%
where $f(r)$\ and $R(r)$\ are given by%
\begin{equation}
f(r)=-\frac{\alpha ^{2}+1}{\alpha ^{2}-1}\left( \frac{b}{r}\right)
^{-2\gamma }-\frac{m}{r^{1-2\gamma }}+\frac{\left( \alpha ^{2}+1\right)
^{2}r^{2}}{\alpha ^{2}-3}\left( \frac{b}{r}\right) ^{2\gamma }\left\{
\Lambda +2\beta ^{2}\left[ _{2}\mathcal{F}_{1}\left( -\frac{1}{2},\frac{%
\alpha ^{2}-3}{4},\frac{\alpha ^{2}+1}{4},-\eta \right) -1\right] \right\} ,
\end{equation}%
\begin{equation}
R(r)=e^{\alpha \Phi },
\end{equation}%
in which $\alpha $\ is an arbitrary dilaton coupling constant determining
the strength of coupling of the scalar and electromagnetic field, $b$ is an
arbitrary constant, $\beta $\ is nonlinearity (Born-Infeld) parameter, and $%
_{2}\mathcal{F}_{1}$ is a hypergeometric function. In addition, $\gamma =%
\frac{\alpha ^{2}}{\alpha ^{2}+1}$, $\eta =\frac{q^{2}}{\beta ^{2}r^{4}}%
\left( \frac{r}{b}\right) ^{4\gamma }$,\ and $\Phi $ is the dilaton field%
\begin{equation}
\Phi (r)=\frac{\gamma }{\alpha }\ln \left( \frac{b}{r}\right) .
\end{equation}

Using the definition of the surface gravity, one can obtain the Hawking
temperature of the black hole on the outermost horizon, $r_{+}$,%
\begin{equation}
T=\frac{\alpha ^{2}+1}{2\pi r_{+}\left( \alpha ^{2}-1\right) }\left( \frac{b%
}{r_{+}}\right) ^{2\gamma }\left[ r_{+}^{2}\left( \alpha ^{2}-1\right)
\left( \beta ^{2}-\frac{\Lambda }{2}\right) -\frac{1}{2}\left( \frac{r_{+}}{b%
}\right) ^{4\gamma }-\beta ^{2}r_{+}^{2}\left( \alpha ^{2}-1\right) \sqrt{%
1+\eta _{+}}\right] ,  \label{T}
\end{equation}%
where $\eta _{+}=\left. \eta \right\vert _{r=r_{+}}$. In addition, the
entropy of the black hole per unit volume $V_{2}$\ in Einstein gravity is a
quarter of the event horizon area%
\begin{equation}
S=\frac{A}{4}=\frac{r_{+}^{2}}{4}\left( \frac{b}{r_{+}}\right) ^{2\gamma }.
\label{S}
\end{equation}

The electric potential $U$, measured at infinity with respect to the horizon
is given by the following explicit form%
\begin{equation}
U=\frac{q}{r}\ _{2}\mathcal{F}_{1}\left( \frac{1}{2},\frac{\alpha ^{2}+1}{4},%
\frac{\alpha ^{2}+5}{4},-\eta _{+}\right) ,
\end{equation}%
where $q$\ is an integration constant which is related to the electric
charge of the black hole. One can use the flux of the electric field at
infinity to obtain the electric charge per unit volume $V_{2}$%
\begin{equation}
Q=\frac{q}{4\pi }.
\end{equation}

The total mass of obtained black holes per unit volume $V_{2}$ can be
obtained by using the behavior of the metric at large $r$
\begin{equation}
M=\frac{mb^{2\gamma }}{8\pi \left( \alpha ^{2}+1\right) }.  \label{M}
\end{equation}

It has been shown that by considering the entropy and electric charge as a
complete set of extensive parameters, these conserved and thermodynamical
quantities satisfy the first law of thermodynamics \cite{Sheykhi}%
\begin{equation}
dM=TdS+UdQ.
\end{equation}

\section{REENTRANT PHASE TRANSITION \label{PV}}

In the extended phase space, the negative cosmological constant is
considered as a positive thermodynamical pressure \cite{Caldarelli,Kastor},
which in our case it has the following form
\begin{equation}
P=-\frac{\Lambda }{8\pi }\left( \frac{b}{r_{+}}\right) ^{2\gamma }.
\label{P1}
\end{equation}

It is worthwhile to mention that in order to find the relation
between the cosmological constant and the pressure, one can use
the components of energy-momentum tensor in diagonal form.
Regarding the energy-momentum tensor, we find that although in the
absence of dilaton field the traditional definition of pressure
($P=-\Lambda /(8\pi )$) is obtained, such definition may be
generalized in the presence of dilaton field. It is worth
mentioning that if one uses the traditional relation of the
pressure in the presence of dilaton gravity, the same van der
Waals like phase transition may be seen, but with different
critical quantities, which is not correct.

On the other hand, in the absence of dilaton field ($\gamma =0$), the
pressure (\ref{P1}) takes the standard form (\ref{P2}). In this situation,
the total mass (\ref{M}) plays the role of enthalpy of system \cite{Kastor},
and the Smarr formula is given by
\begin{equation}
M=2\left( 1-\gamma \right) TS+UQ+\left( 2\gamma -1\right) \left( 2VP+%
\mathcal{B}\beta \right) ;\ \ \ \ \ \mathcal{B}=\left( \frac{\partial M}{%
\partial \beta }\right) _{S,Q,P},  \label{Smarr}
\end{equation}

The validity of the first law of thermodynamics in the extended phase space
should be written as a differential equation for the enthalpy with the
following form
\begin{equation}
dM=TdS+UdQ+VdP+\mathcal{B}d\beta ,  \label{FL}
\end{equation}%
where $V$ is the thermodynamical volume conjugate to $P$
\begin{equation}
V=\left( \frac{\partial M}{\partial P}\right) _{S,Q,\beta }=\frac{\alpha
^{2}+1}{3-\alpha ^{2}}b^{2\gamma }r_{+}^{(\alpha ^{2}+3)/(\alpha ^{2}+1)}.
\end{equation}
\begin{figure}[tbp]
$%
\begin{array}{ccc}
\epsfxsize=8.5cm \epsffile{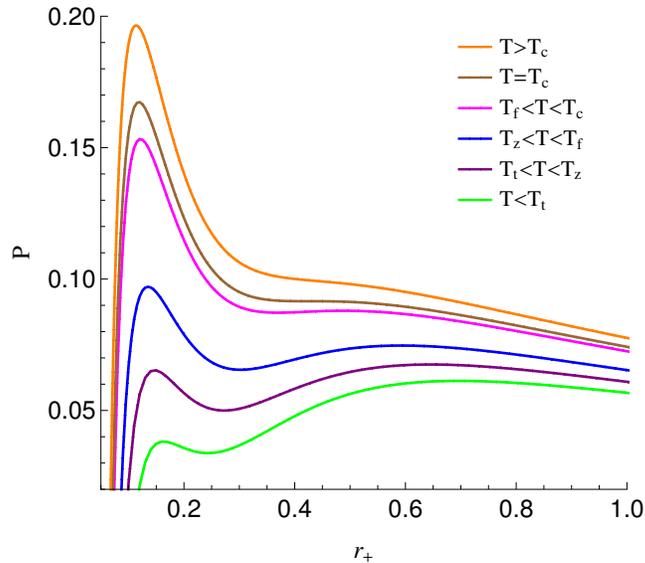} &  &
\end{array}
$%
\caption{$P-r_{+}$ diagram for different regions of temperature. At $T=T_{c}$
there is an inflection point which shows a second order phase transition
between SBH and LBH, and black holes are physically indistinguishable for $%
T>T_{c}$. For $T_{f}<T<T_{c}$ ($T_{z}<T<T_{f}$), black holes undergo a
zeroth (first) order SBH-LBH phase transition. The RPT is located at $%
T_{t}<T<T_{z}$ and black holes undergo the LBH-SBH-IBH phase transition in
this region of temperature. There are LBH at $T<T_{t}$ and there is no phase
transition in this region.}
\label{PVfig}
\end{figure}
\begin{figure}[tbp]
$%
\begin{array}{ccc}
\epsfxsize=7.5cm \epsffile{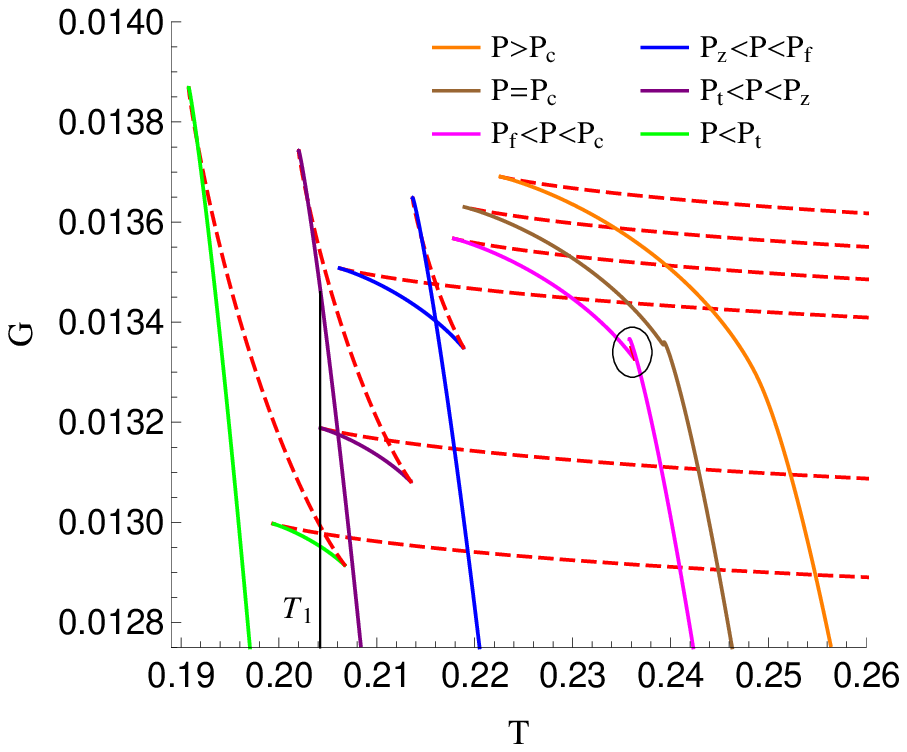} & \epsfxsize=7.5cm \epsffile{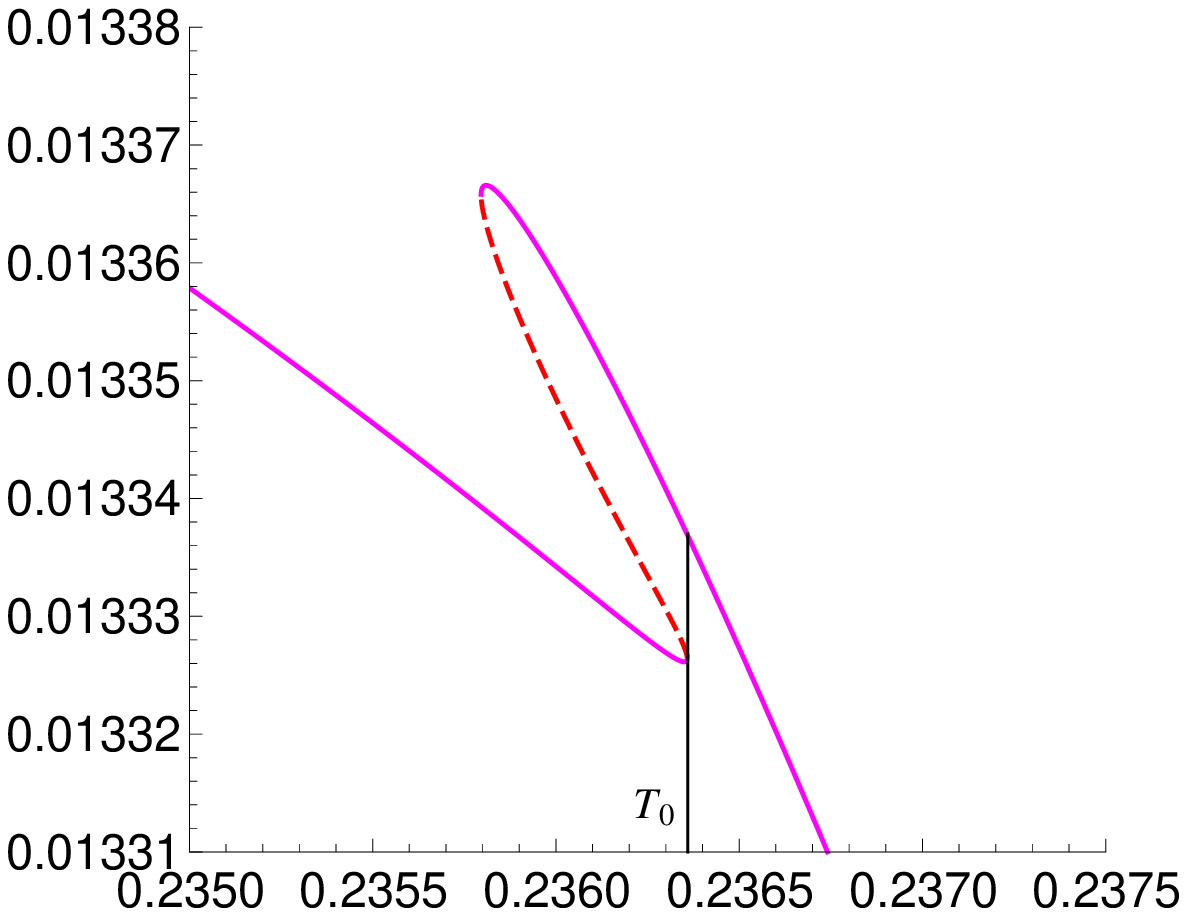} &
\end{array}
$%
\caption{$G-T$ diagram for different regions of pressure. The right panel
indicates a close-up of the zeroth order SBH-LBH phase transition located at
the small circle in the left panel. The solid lines correspond to $C_{P}>0$
whereas the dashed red lines are related to $C_{P}<0$. At the joins of
dashed and solid lines, $C_{P}$ diverges. In addition, $G-T$ curves are
shifted for clarity. The vertical line at $T_{1}\in \left(
T_{t},T_{z}\right) $ ($T_{0}\in \left( T_{f},T_{c}\right) $) indicates a
discontinuity in the Gibbs free energy which shows a zeroth order SBH-IBH
(SBH-LBH) phase transition.}
\label{GTfig}
\end{figure}
\begin{figure}[tbp]
$%
\begin{array}{ccc}
\epsfxsize=7.5cm \epsffile{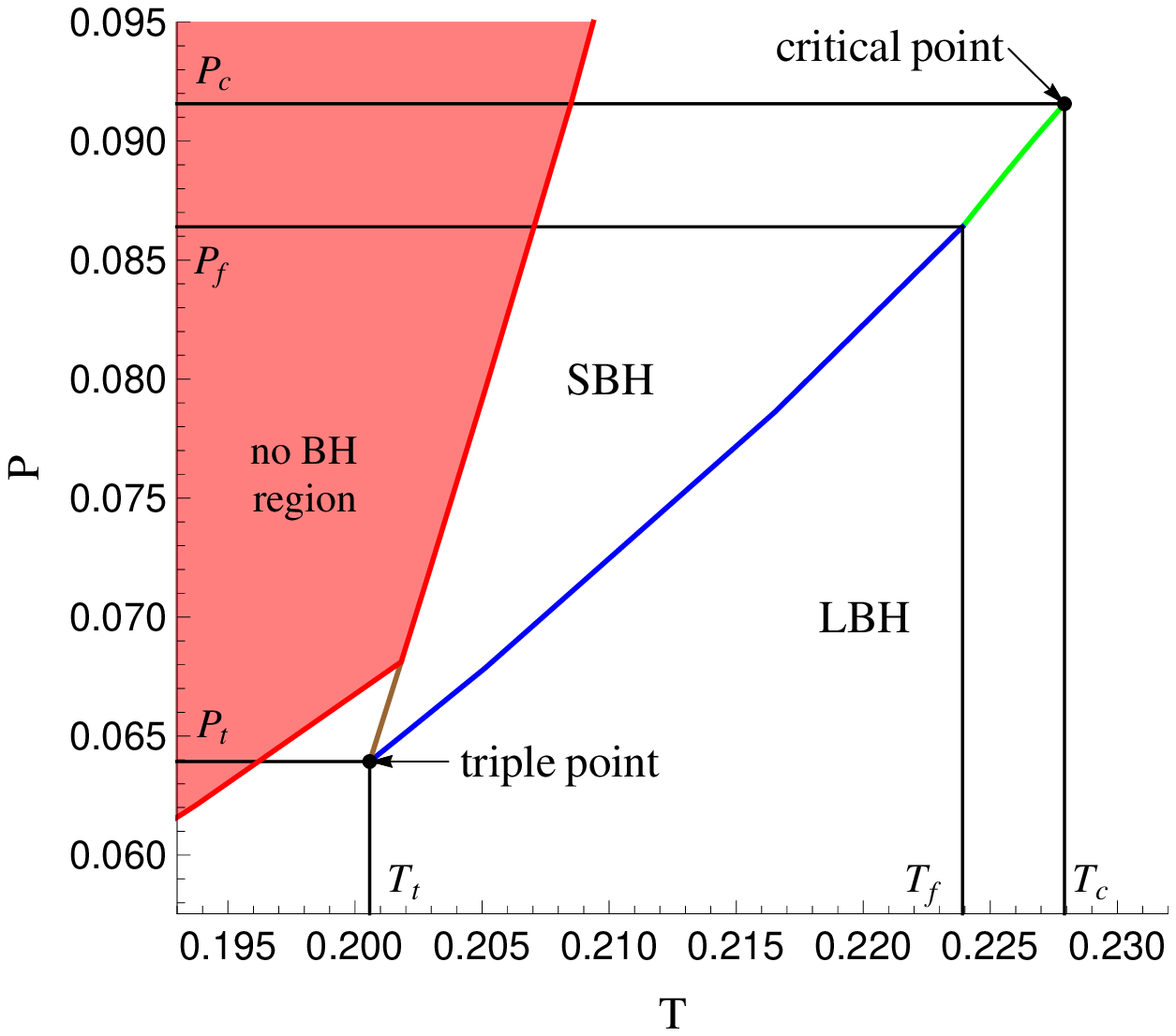} & \epsfxsize=7.5cm \epsffile{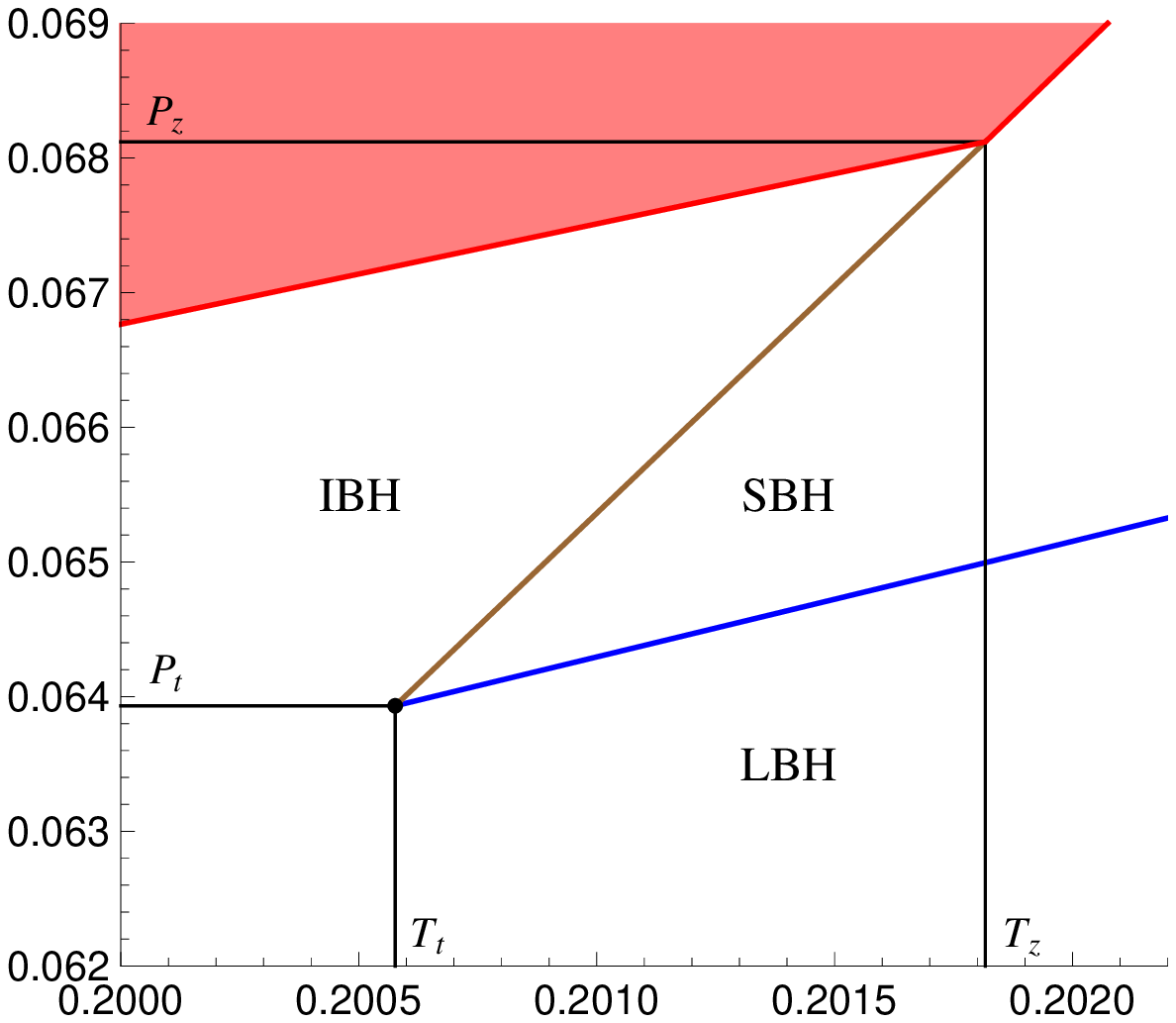} &
\end{array}
$%
\caption{$P-T$ diagram. The shaded area illustrates the no black hole region
and the right panel indicates a close-up of the RPT region. The blue and
green curves indicate the coexistence lines of SBH and LBH, whereas the
brown curve corresponds to the coexistence line of IBH and SBH. On crossing
the blue (brown) line, the system goes under a first (zeroth) order phase
transition between SBH and LBH (SBH and IBH). In addition, on crossing the
green line, the system undergoes a zeroth order SBH-LBH phase transition.}
\label{PTfig}
\end{figure}
On the other hand, although it is possible to add some additional
terms like $Bdb$ ($B=\partial M/\partial b$) and $Ad\alpha $
($A=\partial M/\partial \alpha $) to the (differential form of)
first law, mathematically, we are not allowed to do this. It is
expected that all intensive and extensive thermodynamic parameters
appear in the Smarr formula (\ref{Smarr}). Therefore, we
considered $b$ and $\alpha $ as two constants (not thermodynamic
variables) since they do not appeared in the Smarr formula.

Here, we study the thermodynamics of $4$-dimensional Born-Infeld-dilaton
black holes in the canonical ensemble (fixed $Q$ and $\beta $) of extended
phase space. So, by using the temperature (\ref{T}) and the relation between
the cosmological constant and pressure (\ref{P1}), it is straightforward to
show that the equation of state, $P=P\left( r_{+},T\right) $,\ is given by%
\begin{equation}
P=\frac{T}{2r_{+}\left( \alpha ^{2}+1\right) }+\frac{1}{8\pi r_{+}^{2}}\left[
\frac{1}{\left( \alpha ^{2}-1\right) }\left( \frac{b}{r_{+}}\right)
^{-2\gamma }-2\beta ^{2}r_{+}^{2}\left( \frac{b}{r_{+}}\right) ^{2\gamma
}\left( 1-\sqrt{1+\eta _{+}}\right) \right] .  \label{P}
\end{equation}

The thermodynamical behavior of the system is governed by the Gibbs free
energy, therefore, we should obtain the Gibbs free energy as well. In the
extended phase space, one can determine the Gibbs free energy per unit
volume $V_{2}$ by using the following definition
\begin{eqnarray}
G &=&M-TS=\frac{1}{16\pi }\left\{ 2\beta ^{2}r_{+}^{3}\left( \alpha
^{2}+1\right) \left( \frac{b}{r_{+}}\right) ^{4\gamma }\left( \sqrt{1+\eta
_{+}}-1\right) \right.  \notag \\
&&\left. -\frac{4\left( \alpha ^{2}+1\right) }{\left( \alpha ^{2}-3\right) }%
b^{4\gamma }r_{+}^{(3-\alpha ^{2})/(\alpha ^{2}+1)}\left[ 4\pi P\left( \frac{%
b}{r_{+}}\right) ^{-2\gamma }+\beta ^{2}\left( 1-\mathcal{H}\right) \right]
-r_{+}\left[ 8\pi r_{+}^{2}P\left( \alpha ^{2}+1\right) \left( \frac{b}{r_{+}%
}\right) ^{2\gamma }-1\right] \right\} ,  \label{GG}
\end{eqnarray}%
where $\mathcal{H}=\ _{2}\mathcal{F}_{1}\left( -\frac{1}{2},\frac{\alpha
^{2}-3}{4},\frac{\alpha ^{2}+1}{4},-\eta _{+}\right) $.

On the other hand, the heat capacity in extended phase space at constant
pressure is%
\begin{equation}
C_{P}=T\left( \frac{\partial S}{\partial T}\right) _{P},  \label{Cp}
\end{equation}%
which can be obtained easily by using Eqs. (\ref{T})\ and (\ref{S}). It is
worthwhile to mention that the heat capacity at constant $P$\ is, in fact, a
heat capacity at constant $P$, $Q$, and $\beta $ because we are working in
the canonical ensemble. The negativity of $C_{P}$\ represents unstable black
holes, whereas its positivity indicates stable ones. In order to study the
phase transition of black holes, one can use the definition of inflection
point%
\begin{equation}
\left. \frac{\partial P(r_{+},T)}{\partial r_{+}}\right\vert
_{T=T_{c},r_{+}=r_{+c}}=\left. \frac{\partial ^{2}P(r_{+},T)}{\partial
r_{+}^{2}}\right\vert _{T=T_{c},r_{+}=r_{+c}}=0,  \label{IF}
\end{equation}%
which can be used to obtain the critical horizon radius $r_{+c}$ and
temperature $T_{c}$. Due to the complexity of obtained relation, we cannot
calculate $r_{+c}$ analytically. So, we use the numerical method and some
diagrams to investigate the RPT of the black holes for the fixed values of $%
q=0.2$, $\alpha =0.1$, $\beta =2$, and $b=1$.

The general behavior of Born-Infeld-dilaton black holes is illustrated in
Figs. \ref{PVfig}, \ref{GTfig}, and \ref{PTfig}. Fig. \ref{PVfig} (\ref%
{GTfig}) has been plotted for different regions of temperature (pressure) in
$P-r_{+}$\ ($G-T$) diagram, and Fig. \ref{PTfig} shows the coexistence
lines. The different areas of temperature and\ pressure in Figs. \ref{PVfig}
and \ref{GTfig} are equivalent to each other. For $P>P_{c}$ in $G-T$
diagram, the curve looks like the Hawking-Page phase transition \cite{HP}.
The dashed red line describes small unstable black holes with the negative
heat capacity whereas the solid line corresponds to stable large black holes
with the positive heat capacity (for more discussion about the relation
between the heat capacity and Gibbs energy see \cite{MassiveYM}).
Considering Figs. \ref{PVfig} and \ref{GTfig}, one can see that there is a
critical point at $P=P_{c}$ in $G-T$ diagram (at $T=T_{c}$ in $P-r_{+}$
diagram which characterized by an inflection point) which shows a second
order phase transition between SBH and LBH. For $P_{f}<P<P_{c}$\ ($%
T_{f}<T<T_{c}$), there is a region in which black holes undergo a zeroth
order SBH-LBH phase transition. The vertical line at $T=T_{0}\in \left(
T_{f},T_{c}\right) $ in Fig. \ref{GTfig} shows this behavior with a finite
jump in the Gibbs energy. This kind of phase transition has never seen
before in the RPT of black holes, and it is due to the presence of the
dilaton field (for more details see table $I$ and related discussion). The
standard first order SBH-LBH phase transition occurs for $P_{z}<P<P_{f}$\ ($%
T_{z}<T<T_{f}$). For $P\in \left( P_{t},P_{z}\right) $ and $T\in \left(
T_{t},T_{z}\right) $, there are three different phases of intermediate black
holes (IBH), SBH, and LBH. The vertical line at $T=T_{1}\in \left(
T_{t},T_{z}\right) $ in Fig. \ref{GTfig} shows a discontinuity in the Gibbs
free energy which describes a zeroth order phase transition between SBH and
IBH. There is also a first order SBH-LBH phase transition in this region of
pressures and temperatures. This behavior is known as RPT. It is worthwhile
to mention that IBH are macroscopically similar to LBH. Therefore, black
holes undergo the large-small-large phase transition in this region of
pressures. Finally, there are just LBH for $P<P_{t}$ and $T<T_{t}$.

Fig. \ref{PTfig} shows the coexistence lines of SBH+LBH (the blue and green
lines) and IBH+SBH (the brown line) in different scales. The green line is
bounded by a critical point ($T_{c}$,$P_{c}$) and point ($T_{f}$,$P_{f}$)
between SBH and LBH. Similarly, the blue line is bounded by this point ($%
T_{f}$,$P_{f}$)\ and triple point ($T_{t}$,$P_{t}$) between SBH and LBH. In
addition, the brown line is bounded by this triple point ($T_{t}$,$P_{t}$)
and point ($T_{z}$,$P_{z}$) between SBH and IBH. When black hole crosses the
blue (brown) line from left to right or top to bottom, it undergoes a first
(zeroth) order phase transition from SBH to LBH (IBH to SBH). So, one can
see the RPT behavior of Born-Infeld-dilaton black holes for a narrow range
of temperatures $T\in \left( T_{t},T_{z}\right) $\ and pressures $P\in
\left( P_{t},P_{z}\right) $. On the other hand, a zeroth order phase
transition from SBH to LBH does occur whenever the black hole crosses the
green line from left to right or top to bottom.

Table $I$ shows some values for parameters $\alpha $\ and $\beta $\ which
the black hole case study experiences the RPT investigated in this paper. We
generated this table for fixed values of $q$\ and $b$\ due to the fact that
these two parameters do not change the general behavior of the phase
transition structure. From this table, one can find that the modification of
zeroth order SBH-LBH phase transition below the critical point disappear
whenever the dilaton parameter $\alpha $\ is small enough. Thus, the dilaton
field is responsible for this modification. In addition, we find that there
is an $\alpha $, say $\alpha _{1}$, with $0.008<\alpha _{1}<0.02$\ in which
we have the common RPT of black holes for $\alpha <\alpha _{1}$\ and black
holes undergo a zeroth order SBH-LBH phase transition below the critical
point for $\alpha _{1}<\alpha <\alpha _{2}$ (we will obtain this $\alpha
_{2} $\ in coming section).

\begin{center}
\begin{tabular}{ccccccccccccccccccc}
\hline\hline
$\beta $ &  & $\alpha $ &  & $r_{+c}$ &  & $T_{c}$ &  & $P_{c}$ & $%
\left\Vert \frac{{}}{{}}\right. $ & $\alpha $ &  & $\beta $ &  & $r_{+c}$ &
& $T_{c}$ &  & $P_{c}$ \\ \hline\hline
$^{\ast }2.000$ &  & $8.000\times 10^{-3}$ &  & $0.4162$ &  & $0.2294$ &  & $%
0.09516$ & $\left\Vert \frac{{}}{{}}\right. $ & $1.000\times 10^{-1}$ &  & $%
1.850$ &  & $0.4031$ &  & $0.2310$ &  & $0.09473$ \\ \hline
$2.000$ &  & $2.000\times 10^{-2}$ &  & $0.4165$ &  & $0.2293$ &  & $0.09504$
& $\left\Vert \frac{{}}{{}}\right. $ & $1.000\times 10^{-1}$ &  & $1.910$ &
& $0.4132$ &  & $0.2296$ &  & $0.09327$ \\ \hline
$2.000$ &  & $6.000\times 10^{-2}$ &  & $0.4193$ &  & $0.2289$ &  & $0.09384$
& $\left\Vert \frac{{}}{{}}\right. $ & $1.000\times 10^{-1}$ &  & $1.970$ &
& $0.4211$ &  & $0.2284$ &  & $0.09209$ \\ \hline
$2.000$ &  & $1.000\times 10^{-1}$ &  & $0.4245$ &  & $0.2279$ &  & $0.09157$
& $\left\Vert \frac{{}}{{}}\right. $ & $1.000\times 10^{-1}$ &  & $2.060$ &
& $0.4304$ &  & $0.2270$ &  & $0.09065$ \\ \hline
$2.000$ &  & $1.200\times 10^{-1}$ &  & $0.4278$ &  & $0.2273$ &  & $0.09008$
& $\left\Vert \frac{{}}{{}}\right. $ & $1.000\times 10^{-1}$ &  & $2.120$ &
& $0.4354$ &  & $0.2262$ &  & $0.08987$ \\ \hline
$2.000$ &  & $1.400\times 10^{-1}$ &  & $0.4314$ &  & $0.2267$ &  & $0.08840$
& $\left\Vert \frac{{}}{{}}\right. $ & $1.000\times 10^{-1}$ &  & $2.180$ &
& $0.4397$ &  & $0.2255$ &  & $0.08918$ \\ \hline
$2.000$ &  & $1.500\times 10^{-1}$ &  & $0.4334$ &  & $0.2264$ &  & $0.08749$
& $\left\Vert \frac{{}}{{}}\right. $ & $1.000\times 10^{-1}$ &  & $2.240$ &
& $0.4434$ &  & $0.2249$ &  & $0.08858$ \\ \hline
\end{tabular}%
\\[0pt]

Table $I$: The critical values for RPT for $q=0.2$ and $b=1$. The row
specified with $\ast $ shows the common RPT of black holes in which there is
no zeroth order SBH-LBH phase transition below the critical point. The
behavior of the system for the other rows is just like the case investigated
throughout this paper.
\end{center}

As a final remark of this section, we should note that based on our
numerical analysis and plotted figures, and also the results of other
related references, it seems that regardless of mass and electric charge,
other hairs of a black hole (such as rotation, dilaton scalar field and etc)
have considerable role for the existence of RPT.

\section{zeroth order sbh-lbh phase transition \label{ZO}}

\begin{figure}[tbp]
$%
\begin{array}{ccc}
\epsfxsize=7.5cm \epsffile{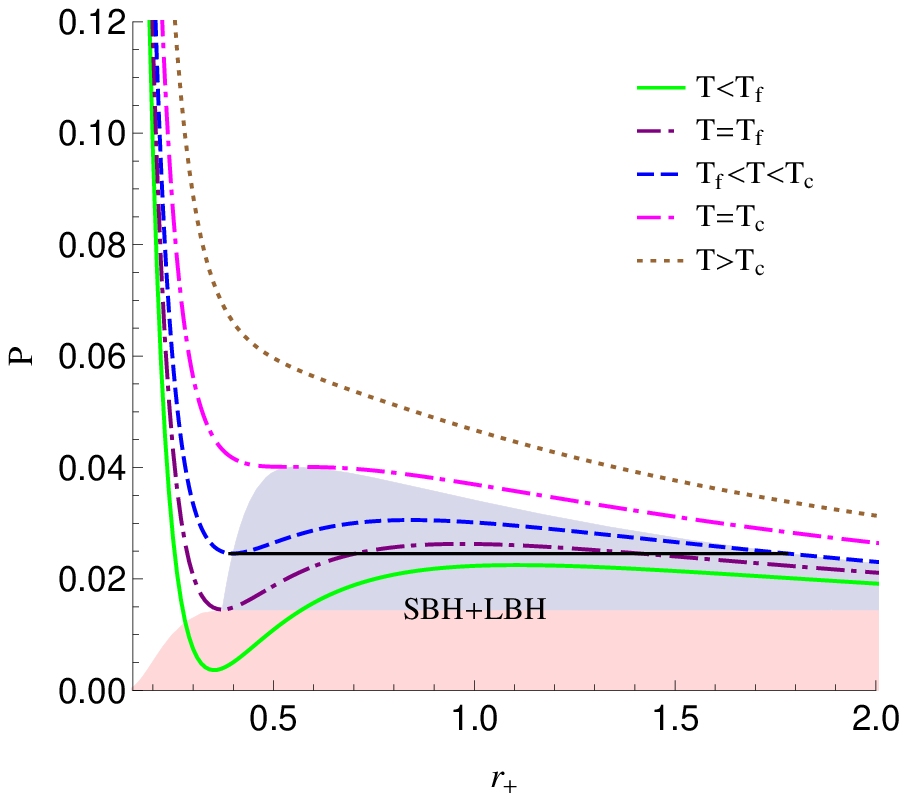} & \epsfxsize=7.5cm \epsffile{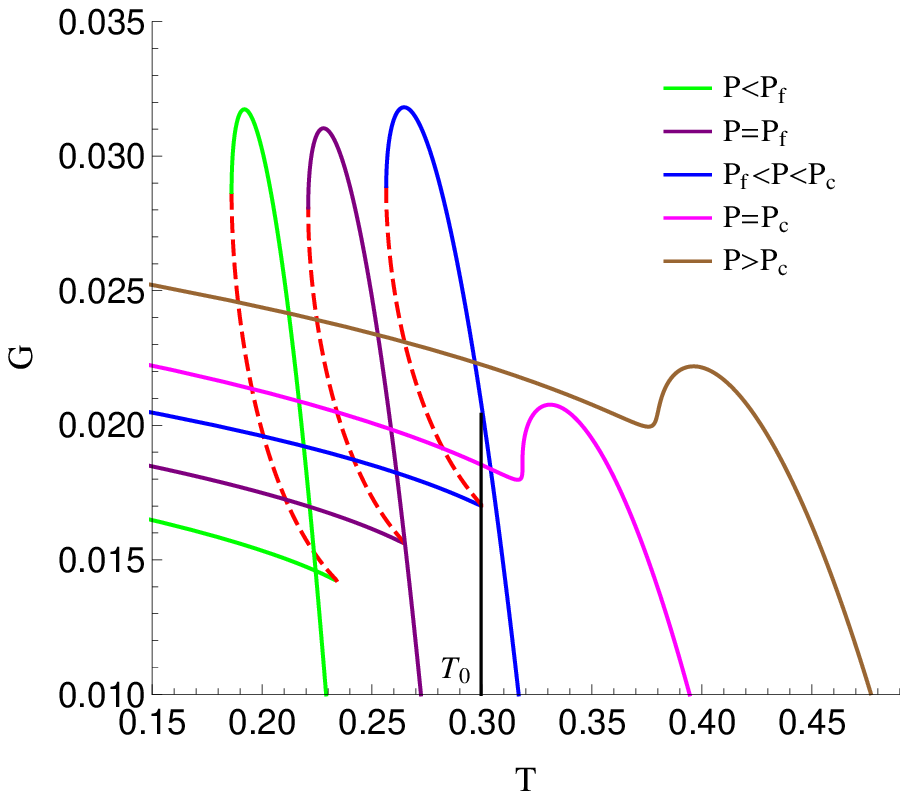}
&
\end{array}
$%
\caption{$P-r_{+}$ and $G-T$ diagrams for different regions of temperature
and pressure. In $P-r_{+}$ diagram, the horizontal black line indicates a
zeroth order SBH-LBH phase transition. Besides, the shaded area indicates
SBH+LBH coexistence region. In the right panel, the solid lines correspond
to $C_{P}>0$ whereas the dashed red lines are related to $C_{P}<0$. At the
joins of dashed and solid lines, $C_{P}$ diverges. In addition, $G-T$ curves
are shifted for clarity. The vertical line at $T_{0}\in \left(
T_{f},T_{c}\right) $, indicates a finite jump in the Gibbs energy which
shows a zeroth order SBH-LBH phase transition.}
\label{PVzero}
\end{figure}
\begin{figure}[tbp]
$%
\begin{array}{ccc}
\epsfxsize=8.5cm \epsffile{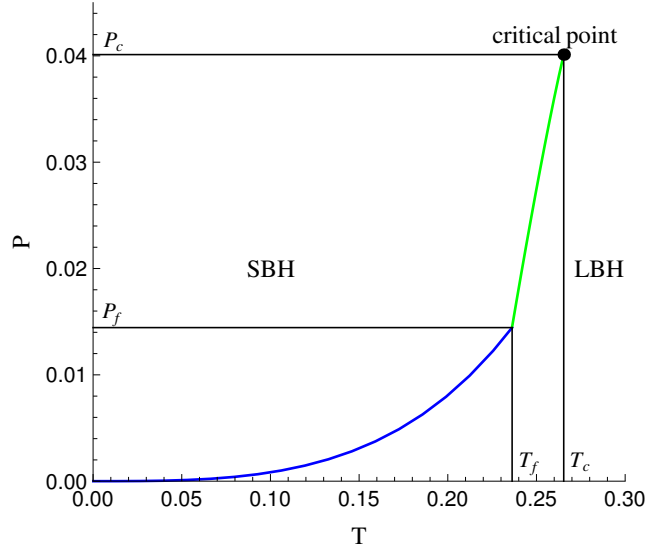} &  &
\end{array}
$%
\caption{$P-T$ diagram. The blue and green curves indicate the coexistence
line of SBH and LBH. On crossing the blue (green) line from left to right or
top to bottom, the system goes under a first (zeroth) order phase transition
from SBH to LBH.}
\label{PTzero}
\end{figure}

In this section, we are going to show that our black hole case study can
undergo a zeroth order SBH-LBH phase transition by adopting suitable
parameters. This kind of phase transition is characterized by a finite jump
in Gibbs energy similar to the one that has been shown in the last section.
Note that this zeroth order phase transition is different from the previous
one and it occurs between SBH and LBH. Such a phase transition has been
obtained for dilaton black holes in \cite{Dehyadegari} and it modifies the
standard first order SBH-LBH phase transition.

In analogy manner with Ref. \cite{Dayyani}, here, we use the fixed values of
$q=0.2$, $\alpha =0.6$, $\beta =2$, and $b=1$ to plot Figs. \ref{PVzero} and %
\ref{PTzero}. The left panel of Fig. \ref{PVzero} shows isotherm curves of
the system for different regions of temperature, and also, $G-T$\ diagram of
the system for different regions of pressure is illustrated in the right
panel which these areas are equivalent to each other. For $T<T_{f}$ ($%
P<P_{f} $), the system undergoes the standard first order SBH-LBH phase
transition which is similar to a van der Waals fluid. $T_{f}$ ($P_{f}$) is
an upper bound for this behavior and the type of phase transition will
change for $T_{f}<T<T_{c}$\ ($P_{f}<P<P_{c}$). In this region, the order of
phase transition will change into zeroth order. Finally, black holes undergo
a second order SBH-LBH phase transition at the critical point ($T_{c}$,$%
P_{c} $), and at temperatures and pressures above $T_{c}$ and $P_{c}$, SBH
and LBH are physically indistinguishable. Therefore, the zeroth order
SBH-LBH phase transition is located between the standard first order phase
transition and critical point (see Fig. \ref{PTzero} for more details).

From Fig. \ref{PTzero}, we find that the coexistence line of first
order phase transition is nonlinear whereas the zeroth order
segment looks linear. However, accurate numeric analysis of the
zeroth order phase transition shows that the related line segment
in the $P-T$ diagram is not actually linear and our data show
nonlinear behavior. This line segment is just like a very small
part of a circle which looks as a line. On the other hand, the
slope of the coexistence line represents the ratio of changes in
entropy and volume, $\Delta S/\Delta r_{+}=\left(
S_{L}-S_{S}\right) /\left( r_{+L}-r_{+S}\right) $, where the
subscripts $L$ and $S$ stand for LBH and SBH, respectively.
Therefore, we find that at the first order phase transition, the
ratio $\Delta S/\Delta r_{+}$ increases along the coexistence line
when the temperature and pressure increase. But for the zeroth
order case, the value of this ratio changes very slowly.

\begin{center}
\begin{tabular}{ccccccccccccccccccc}
\hline\hline
$\beta $ &  & $\alpha $ &  & $r_{+c}$ &  & $T_{c}$ &  & $P_{c}$ & $%
\left\Vert \frac{{}}{{}}\right. $ & $\alpha $ &  & $\beta $ &  &
$r_{+c}$ & & $T_{c}$ &  & $P_{c}$ \\ \hline\hline $^{\dagger
}2.000$ &  & $1.500\times 10^{-1}$ &  & $0.4334$ &  & $0.2264$ & &
$0.08749$ & $\left\Vert \frac{{}}{{}}\right. $ & $6.000\times
10^{-1}$ & & $1.850$ &  & $0.5444$ &  & $0.2657$ &  & $0.04028$ \\
\hline $2.000$ &  & $2.000\times 10^{-1}$ &  & $0.4440$ &  &
$0.2248$ &  & $0.08244$ & $\left\Vert \frac{{}}{{}}\right. $ &
$6.000\times 10^{-1}$ &  & $1.910$ & & $0.5456$ &  & $0.2655$ &  &
$0.04021$ \\ \hline $2.000$ &  & $4.000\times 10^{-1}$ &  &
$0.4928$ &  & $0.2266$ &  & $0.05933$ & $\left\Vert
\frac{{}}{{}}\right. $ & $6.000\times 10^{-1}$ &  & $1.970$ & &
$0.5467$ &  & $0.2654$ &  & $0.04015$ \\ \hline $2.000$ &  &
$6.000\times 10^{-1}$ &  & $0.5472$ &  & $0.2654$ &  & $0.04012$ &
$\left\Vert \frac{{}}{{}}\right. $ & $6.000\times 10^{-1}$ &  &
$2.060$ & & $0.5482$ &  & $0.2653$ &  & $0.04006$ \\ \hline
$2.000$ &  & $9.000\times 10^{-1}$ &  & $0.6471$ &  & $0.8326$ &
& $0.02279$ & $\left\Vert \frac{{}}{{}}\right. $ & $6.000\times
10^{-1}$ &  & $2.120$ & & $0.5490$ &  & $0.2652$ &  & $0.04002$ \\
\hline $^{\ddagger }2.000$ &  & $9.900\times 10^{-1}$ &  &
$0.6827$ &  & $7.988$ & & $0.01956$ & $\left\Vert
\frac{{}}{{}}\right. $ & $6.000\times 10^{-1}$ & & $2.180$ &  &
$0.5498$ &  & $0.2651$ &  & $0.03997$ \\ \hline
$2.000$ &  & $1.100$ &  & $0.7302$ &  & $-0.7725$ &  & $0.01644$ & $%
\left\Vert \frac{{}}{{}}\right. $ & $6.000\times 10^{-1}$ &  & $2.240$ &  & $%
0.5506$ &  & $0.2650$ &  & $0.03993$ \\ \hline
\end{tabular}%
\\[0pt]

Table $II$: The critical values for zeroth order phase transition
for $q=0.2$ and $b=1$. The row specified with $\dagger $ shows the
RPT of black holes
investigated in the previous section. For the row specified with $\ddagger $%
, the black hole does not experience the first order phase
transition and there is just zeroth order SBH-LBH phase
transition. In addition, the value of $\alpha =1$\ is undefined
(see Eq. (\ref{P})) and the critical temperature is negative for
$\alpha >1$. The behavior of the system for the other rows is just
like the case investigated in this section.
\end{center}

Table $II$ shows some values for parameters $\alpha $\ and $\beta $\ which
the black holes experience the zeroth order phase transition investigated in
this section. From this table, we find that the range of $\alpha _{2}$,
introduced in the previous section, with $0.15<\alpha _{2}<0.2$. For $\alpha
_{2}<\alpha <\alpha _{3}$\ ($0.9<\alpha _{3}<1$) black hole experiences a
zeroth order SBH-LBH phase transition for a range of temperature and
pressure. In addition, for $\alpha _{3}<\alpha <1$\ the standard first order
phase transition segment will disappear and black holes just undergo a
zeroth order phase transition. From tables $I$ and $II$, one can see that
the dilaton parameter strongly affects the phase transition behavior of the
system.

\section{Conclusions \label{Conclusions}}

In this paper, we have considered the cosmological constant as
thermodynamical pressure and studied the thermodynamics of $4$-dimensional
Born-Infeld-dilaton black holes in the canonical ensemble of extended phase
space. We have seen that in addition to the standard van der Waals like
phase transition of these black holes \cite{PVdilatonH,PVdilatonSh}, they
can enjoy a novel RPT. It was shown that this behavior happens for a narrow
range of temperatures\ and pressures. In this range of RPT, black holes
undergo a zeroth order IBH-SBH phase transition and first order SBH-LBH
phase transition. In addition, it was shown that there is a region below the
critical point in which the first order phase transition modified to a
zeroth order type. This behavior has never seen before in the RPT of black
holes and we showed that it is due to the presence of the dilaton field.

Moreover, we have shown that by considering suitable values of free
parameters, a portion of the standard first order SBH-LBH phase transition
can modify into the zeroth order SBH-LBH phase transition. It is worthwhile
to mention that numerical calculations show that this kind of phase
transition is due to the presence of the dilaton field. It is interesting to
investigate the microscopic origin of such RPT based on dilaton effect.

As the final remark, we should mention that three different
behaviors have been found for the black hole case study presented
in this paper until now. One is the standard van der Waals like
phase transition presented in \cite{PVdilatonH,PVdilatonSh}, and
two of them are investigated throughout this paper. All these
three behaviors depend on the choice of free parameters, and
therefore, it may be possible to find some other different
behaviors by adopting different values for the free parameters.

\begin{acknowledgements}
The authors wish to thank the anonymous referee for the
constructive comments that enhanced the quality of this paper. We
wish to thank Shiraz University Research Council. This work has
been supported financially by the Research Institute for Astronomy
and Astrophysics of Maragha, Iran.
\end{acknowledgements}

\end{document}